\documentclass[structabstract]{aa}
\pdfoutput=1

\usepackage{txfonts}
\usepackage{natbib}
\usepackage{graphicx}  
\newcommand{\dt}{\mathrm{d}}
\newcommand{\kmps}{\ensuremath{\mathrm{km~s}^{-1}}}

\begin{document}
\title{A 3D radiative transfer framework: \\ VII. Arbitrary velocity fields in the Eulerian frame}
\author{A. M. Seelmann$^1$, P.H. Hauschildt$^1$ \and E. Baron$^{1,2,3}$}
\institute{Hamburger Sternwarte, Gojenbergsweg 112, 21029 Hamburg, Germany \email [aseelmann,yeti]@hs.uni-hamburg.de \and
           Dept.  of Physics and Astronomy, University of Oklahoma, 440 W. Brooks, Rm 100, Norman, OK 73019 USA \email  baron@ou.edu \and
           Computational Research Division, Lawrence Berkeley National Laboratory, MS 50F-1650, 1 Cyclotron Road, CA 94720-8139 USA}
\date{Accepted 17/07/2010}

\abstract{}
{A solution of the radiative-transfer problem in 3D with arbitrary velocity fields in the Eulerian frame is presented. The method is implemented in our 3D radiative transfer framework and used in the PHOENIX/3D code. It is tested by comparison to our well-tested 1D co-moving frame radiative transfer code, where the treatment of a monotonic velocity field is implemented in the Lagrangian frame. The Eulerian formulation does not need much additional memory and is useable on state-of-the-art computers, even large-scale applications with 1000's of wavelength points are feasible.}
{In the Eulerian formulation of the problem, the photon is seen by the atom at a Doppler-shifted wavelength depending on its propagation direction, which leads to a Doppler-shifted absorption and emission. This leads to a different source function and a different $\Lambda^*$ operator in the radiative transfer equations compared to the static case.}
{The results of the Eulerian 3D spherical calculations are compared to our well-tested 1D Lagrangian spherical calculations, the agreement is, up to $v_{max}=1\cdot10^3 \ \kmps$ very good. Test calculation in other geometries are also shown.}{}


\titlerunning{3D radiative transfer framework. VII.}
\authorrunning{A.M. Seelmann, P.H. Hauschildt and E. Baron}

\maketitle

\section{Introduction}
A solution of the 1D radiative transfer problem in arbitrary velocity fields in the Lagrangian frame has been developed by \citet{CMF_four}. \citet{CMF_five} optimized their method by reducing the memory footprint of the algorithm (using domain decomposition), and also introduced a new method which speeds up the formal solution by developing an iterative Gauss-Seidel (GS) type solver where the solution becomes quasi-analytic when the source function is interpolated linearly. However, in a 3D setup the limiting factor is the memory footprint which already stretches the limits of modern supercomputers. The memory requirements of a 3D  calculation in the Lagrangian frame are very high because one has to store additional wavelength information in every volume element (hereafter voxel) for every solid angle, whereas the 3D Eulerian frame calculation needs more computing time in multi-level applications due to the explicit computing of the opacity for every solid angle point.\\

\par In Section \ref{method} we describe the Eulerian formulation of the problem, in Section \ref{application} the comparison to our well-tested 1D code is presented and application examples in other geometrical setups are shown. In Section \ref{problems} we describe expected and discovered limitations with the formalism in the Eulerian frame.

\section{Method}\label{method}
The 3D radiative transfer framework uses the full-characteristics method to solve the radiative transfer equations \citep{threed_one}. The intensity along a characteristic, which are straight lines with given direction $(\theta,\phi)$ in the Eulerian frame, is simply given by
\begin{equation}
  \frac{\dt I}{\dt\tau} = I - S.
\end{equation}
With this definition, the formal solution for a given characteristic $(\theta,\phi)$ can be written as
\begin{eqnarray}
 I(\tau_i) &=& I(\tau_{i-1}) \mathrm{exp}(\tau_{i-1}-\tau_i) + \int_{\tau_{i-1}}^{\tau_i} S(\tau) \mathrm{exp} (\tau-\tau_i) \dt \tau \\
           &\equiv& I_{i-1}(-\Delta\tau_{i-1}) + \Delta I_{i}
\end{eqnarray}

\noindent where i labels the points along the characteristic, $S$ is the source function and $\Delta\tau_i$ is the optical depth, computed, e.g., by using piecewise linear interpolation of the opacity $\chi$ along the characteristic:
\begin{equation}
\Delta \tau_{i-1} = (\chi_{i-1}(\lambda) + \chi_i(\lambda)) |s_{i-1} - s_i| / 2.
\end{equation}
\citet{threed_one} give a more detailed explanation of the general method.
\par Due to the movement of the atom in the Eulerian frame, the atom 'sees' the photon (on the characteristic) at a wavelength shifted according to
\begin{equation} \label{doppler_shift}
 \lambda_{\rm{atom},i} = \lambda_{\rm{observer},i} \cdot (1 + \frac{\vec{e_{\rm{char}}}\cdot \vec{v_i}}{c})
\end{equation}
where $\vec{e_{\rm{char}}}$ is the unit vector in the direction of the characteristic, $c$ the speed of light and $\vec{v}$ the velocity of the atom. This leads to a different opacity seen by the characteristic depending on its direction:
\begin{equation} \label{opac}
 \chi_i(\lambda) = \chi_i(\lambda,\theta,\phi)
\end{equation}
In the case of line transfer, the profile of the line becomes anisotropic in the Eulerian frame:
\begin{equation} \label{profile_formula}
\Phi(\lambda) = \Phi(\lambda,\theta,\phi)
\end{equation}
The solution for the line transfer then proceeds with the Eulerian profile function $\Phi(\lambda,\theta,\phi)$ and following \citet{threed_two}, we obtain 
\begin{equation}
 \bar{J} = \iiint \Phi (\lambda,\mu,\phi) J_{\lambda} \dt\lambda \dt\mu \dt\phi
\end{equation}
and
\begin{equation}
 \bar{\Lambda^*} = \iiint \Phi (\lambda,\mu,\phi) \Lambda^* \dt\lambda \dt\mu \dt\phi
\end{equation}
where $\lambda$ is the wavelength in the observers frame, $\mu=\cos \theta$ and $\phi$ the solid angle under which the voxel is hit by the characteristic.\\
Relativistic corrections are neglected in the Eulerian frame, as we are working on a implementation of the co-moving frame method into our 3D code \citep[][E.~Baron et al., in preparation]{chen}.

\section{Application Examples}\label{application}
A simple two-level-atom approach was used to test the code, the physical atmosphere setup presented in Section \ref{spherical}-\ref{cartesian} is similar to the one used in \citet{threed_four}, remarks about the numerical resolution of our spherical 3D code can also be found there.

%%%%%%%%%%%%% SPHERICAL
\subsection{3D Spherical coordinates} \label{spherical}
Our well-tested Lagrangian 1D spherical code was used to compare its results to the new Eulerian formalisms in the 3D RT code to verify the 3D Eulerian Code.\par
\begin{figure}[t]
 \begin{minipage}{0.5\textwidth}
    \includegraphics[width=\textwidth]{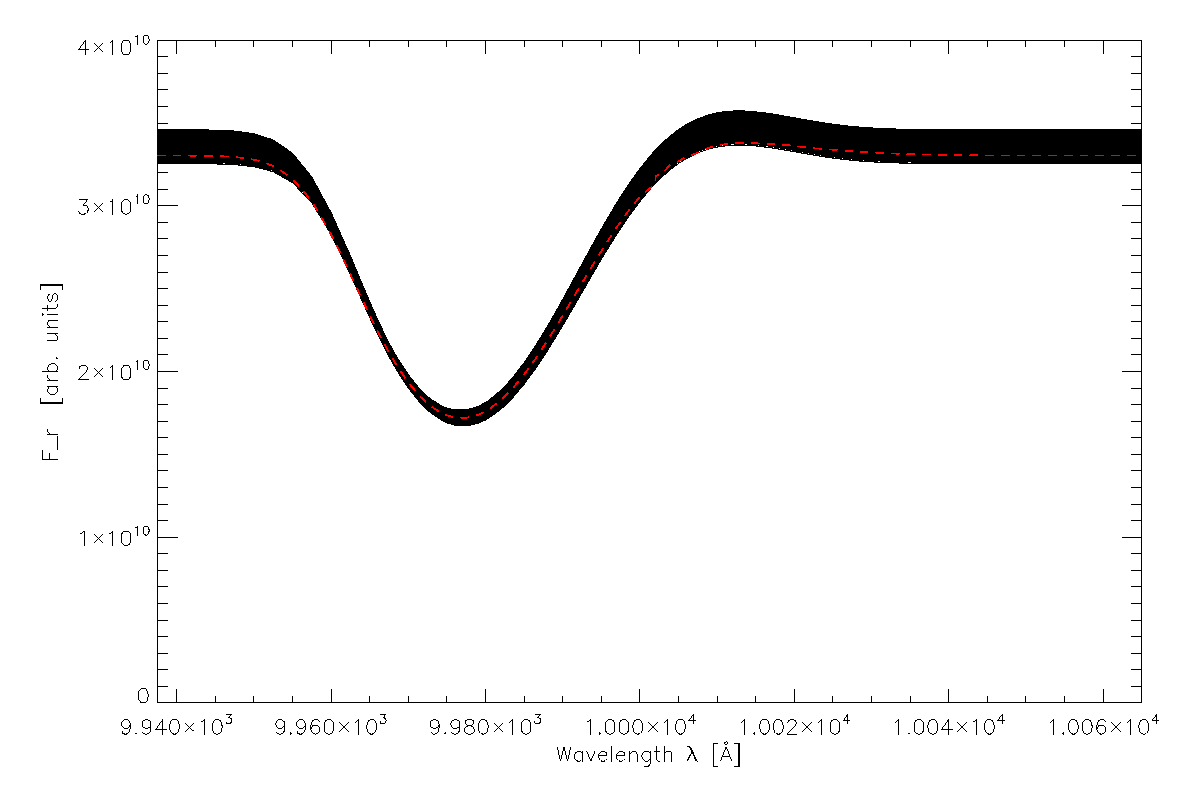} 
  \caption{Comparison of the Eulerian 3D and the Lagrangian 1D code with a linear increasing velocity field with $v_{\rm{max}}=1\cdot10^3\ \kmps$ and scattering $\epsilon_{\rm{line}}=10^{-2}$. Solid: Spectra from all outermost voxels in the 3D spherical coordinate grid, dashed: spectrum from the 1D code.}
  \label{spherical_velo}
 \end{minipage}
\end{figure}

 For this, many test calculations were made with linearly increasing velocity fields up to $v_{\rm{max}}=1\cdot 10^3 \ \kmps$. Figure \ref{spherical_velo} shows the corresponding spectra from a calculation with $v_{\rm{max}}=1\cdot 10^3 \ \kmps$, the agreement to the 1D Lagrangian code is very good, the variation of the 3D lines is due to numerical resolution. The 3D model was computed using $(n_r,n_\theta,n_\phi)=(197x99x197)= 3.842.091$ voxels and a solid angle resolution of $(\theta_c,\phi_c)=(64,64)$ (computing time: ca. 18h on 2048 CPUs) \footnote{Intel Xeon Harpertown CPU's, 92 wavelength points}.

%%%%%%%%%%%%%%%% CARTESIAN
\subsection{3D Cartesian coordinates with and without Periodic Boundary Conditions}\label{cartesian}
\begin{figure}[tb] 
 \begin{minipage}{0.5\textwidth}
   \includegraphics[width=\textwidth]{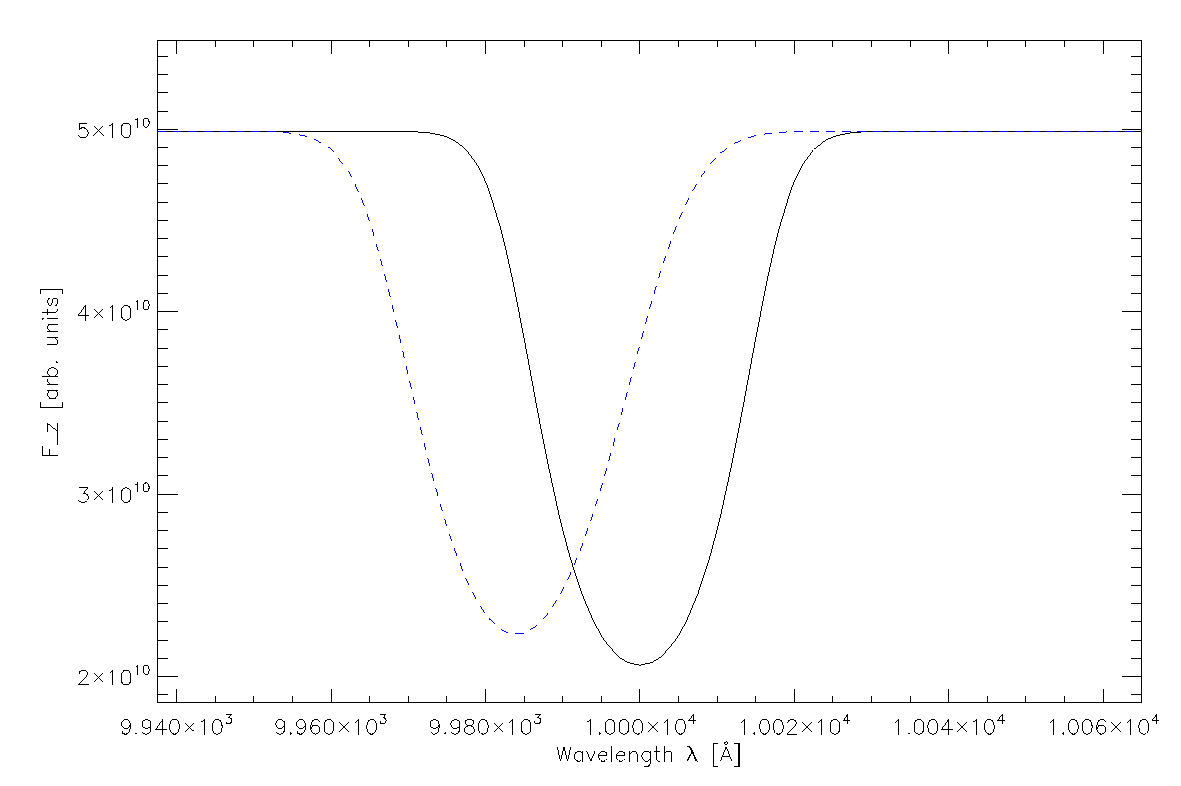} 
  \caption{Spectra from all outermost voxels in the 3D Cartesian geometry with periodic boundary conditions, a linearly increasing velocity field with $v_{max}=1\cdot10^3\ \kmps$ and line scattering $\epsilon_{\rm{line}}=10^{-2}$. Solid: Without the treatment of the velocity field, dashed: with the treatment of the velocity field. This plot shows the influence of the velocity field on the line.}
  \label{pbc_velo}
 \end{minipage}
\end{figure}

In Figure \ref{pbc_velo} a test calculation of the 3D Cartesian code with periodic boundary conditions in a  $(n_x,n_y,n_z)=(157x157x157)=3.869.893$ voxel grid with $(\theta_c,\phi_c)=(64,64)$ is shown (computing time: ca. 6h on 2048 CPUs) $^1$. The difference between the calculation with and without velocity field is clearly visible.  Similar results can be obtained in the Cartesian mode without periodic boundary conditions, the spectra are not shown here.  

%%%%%%%%%%%%%% CYLINDRICAL
%\subsection{3D Cylindrical} \label{cylindrical}
%The method is also implemented in the PHOENIX/3D cylindrical code, a few test calculations were made in order to test the code, but no explicit spectra are shown here. 

%%%%%%%%%%%%% HYDRO STRUCTURE
\subsection{3D Hydro-structure}
\begin{figure*}[ht] 
 \begin{minipage}{0.49\textwidth}
    \includegraphics[width=\textwidth]{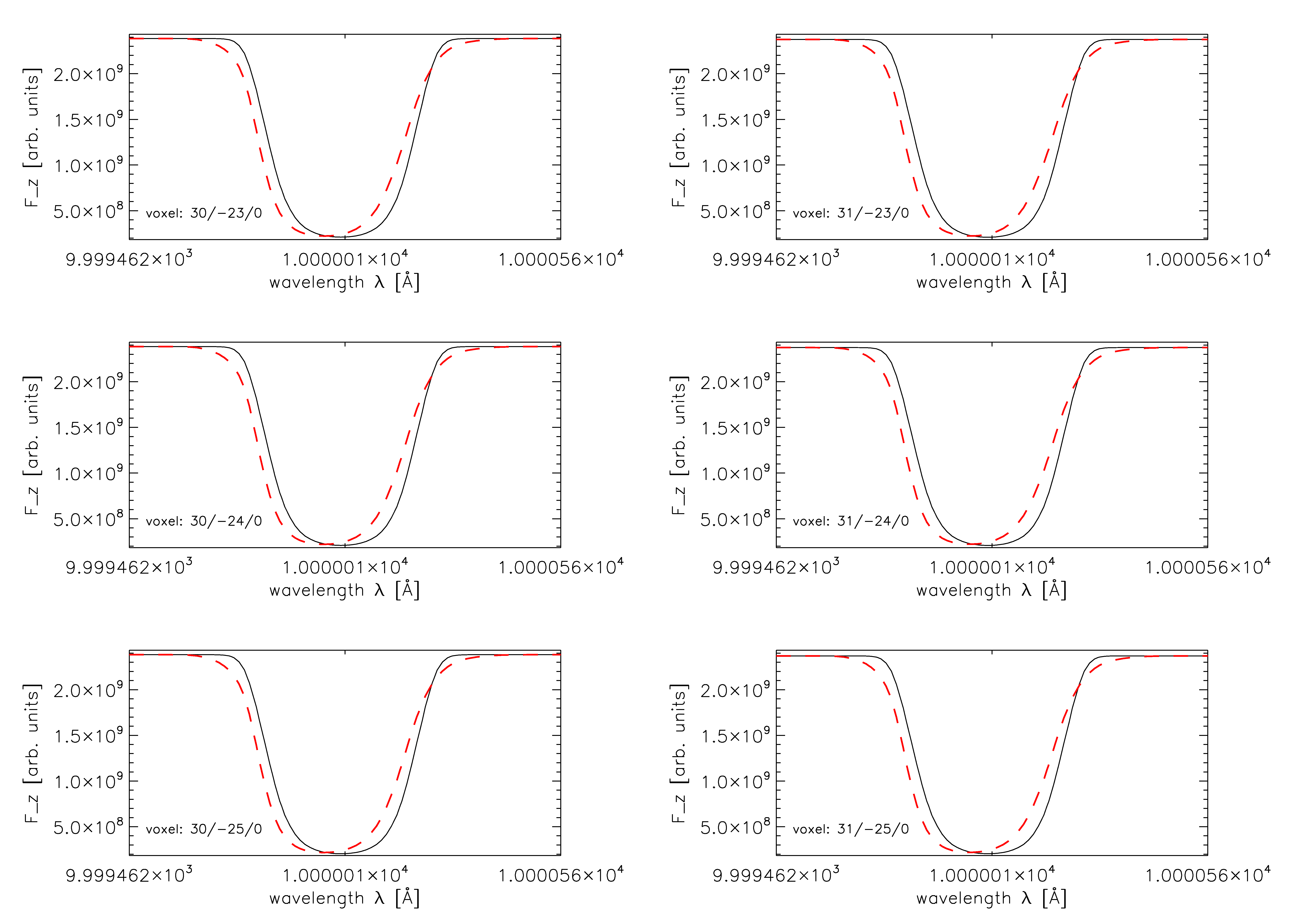} 
  \end{minipage}
\hfill
 \begin{minipage}{0.49\textwidth}
    \includegraphics[width=\textwidth]{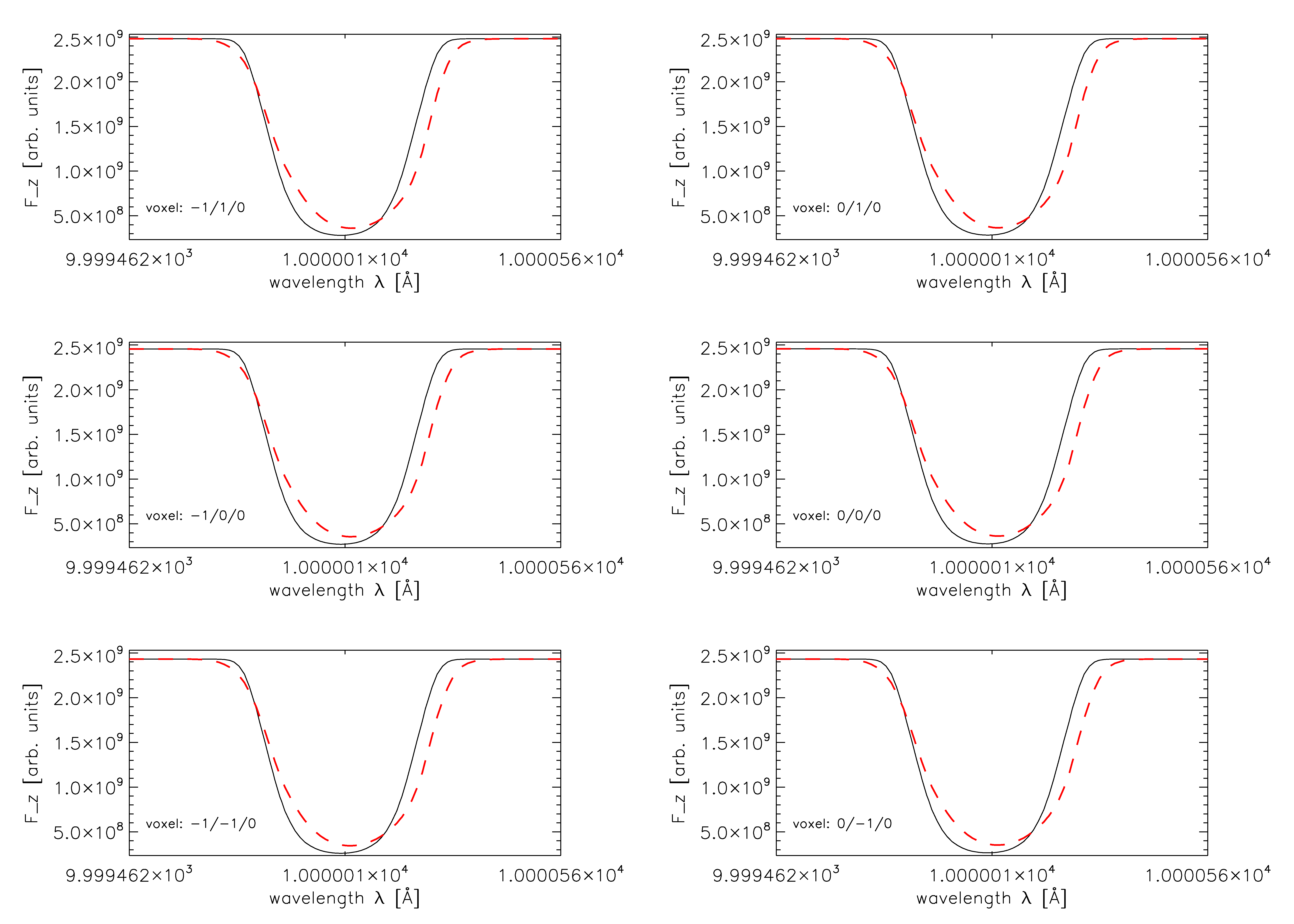} 
  \end{minipage}

  \caption{Spectra from selected voxels on top of the 3D hydro-dynamical Cartesian atmosphere to show the influence of the inherent velocity field on the line, the calculations include line scattering ($\epsilon_{\rm{line}}=10^{-2}$). Solid: without velocity field, dashed: with the inherent velocity field.}
  \label{hydro_velo}
\end{figure*}

To test the new method with a 3D hydro-dynamical structure with an inherent arbitrary velocity field, we obtained a computed snapshot of convection in the solar atmosphere from Ludwig \citep{ludwig,wedemeyer}, which was used as input for the code. The spectra from a few outermost voxels in a calculation ($(n_x,n_y,n_z)=(140x150x140) = 2.940.000$, $(\theta_c,\phi_c)=(64,64)$, computing time: ca. 11h on 1024 CPUs $^1$) with and without velocity field is shown in Figure \ref{hydro_velo}. The plot show that the use of velocity fields in such calculations is mandatory.

%%%%%%%%%%% LIMITATIONS %%%%%%%%%%%%%%%%%%%%%%%%%%%%%%%%%%%%%%%
\section{Limitations of the Eulerian formalisms}\label{problems}
\subsection{Solid Angle points}
\begin{figure} 
 \begin{minipage}{0.5\textwidth}
    \includegraphics[width=\textwidth]{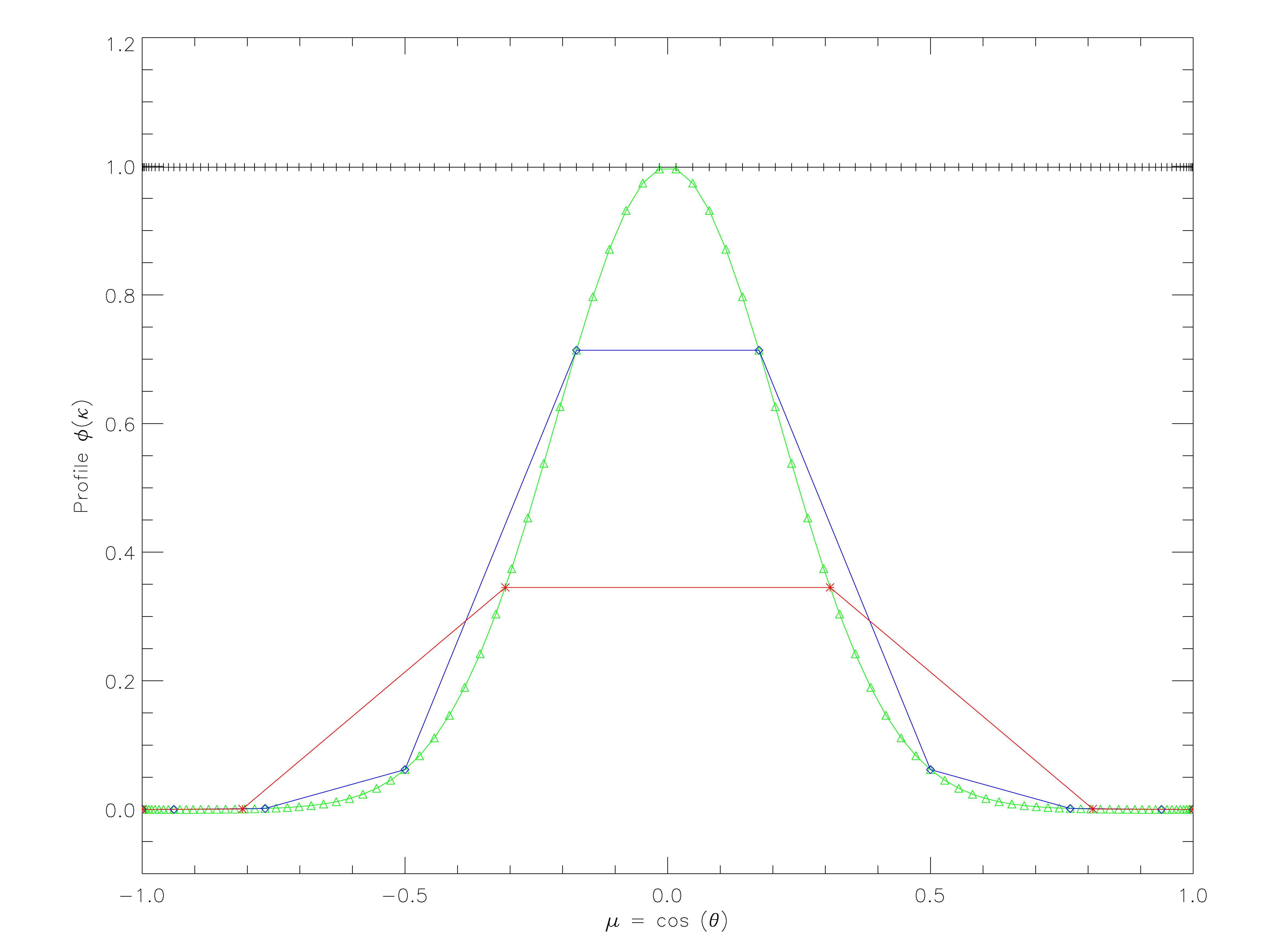} 
  \caption{Plot to illustrate the problem of a poor solid angle discretization. The plus signs show the profile in the line center of a static atmosphere. The asterisks and the diamonds show a poor/medium solid angle discretization, the triangles good solid angle discretization in the Eulerian moving atmosphere where the anisotropic Eulerian profile has  good coverage. See text for a more detailed explanation.} 
  \label{profile_not_hit}
 \end{minipage}
\end{figure}

In static line transfer problems it is necessary that the profile of the line is covered by the discretized wavelength grid used in the calculation, as the profile does only depend on the wavelength $\Phi = \Phi(\lambda)$. In Eulerian moving atmospheres the profile depends on equation \ref{profile_formula} and therefore also on the solid angle discretization.
\\In Figure \ref{profile_not_hit} the profile in the (observer's frame) \textbf{line center} is plotted: The plus signs show the profile in the static case, as it is not solid angle depend it is everywhere $1$. The asterisks show the anisotropic profile of a poor quality, the diamonds of a medium quality solid angle discretization in the moving Eulerian atmosphere: The profile in the line center is not hit at all, this causes a wiggly or even no spectral line. The triangles show a good quality solid angle discretization, the anisotropic profile in the Eulerian moving atmosphere hits the line center and the profile is well covered.
\\Various test show that a solid angle resolution of $(\theta_c,\phi_c)=(64,64)$ is sufficient for velocity fields up to $v_{max}=10^3 \ \kmps$.

\subsection{Relativistic Velocities}
\begin{figure}[h] 
 \begin{minipage}{0.5\textwidth}
    \includegraphics[width=\textwidth]{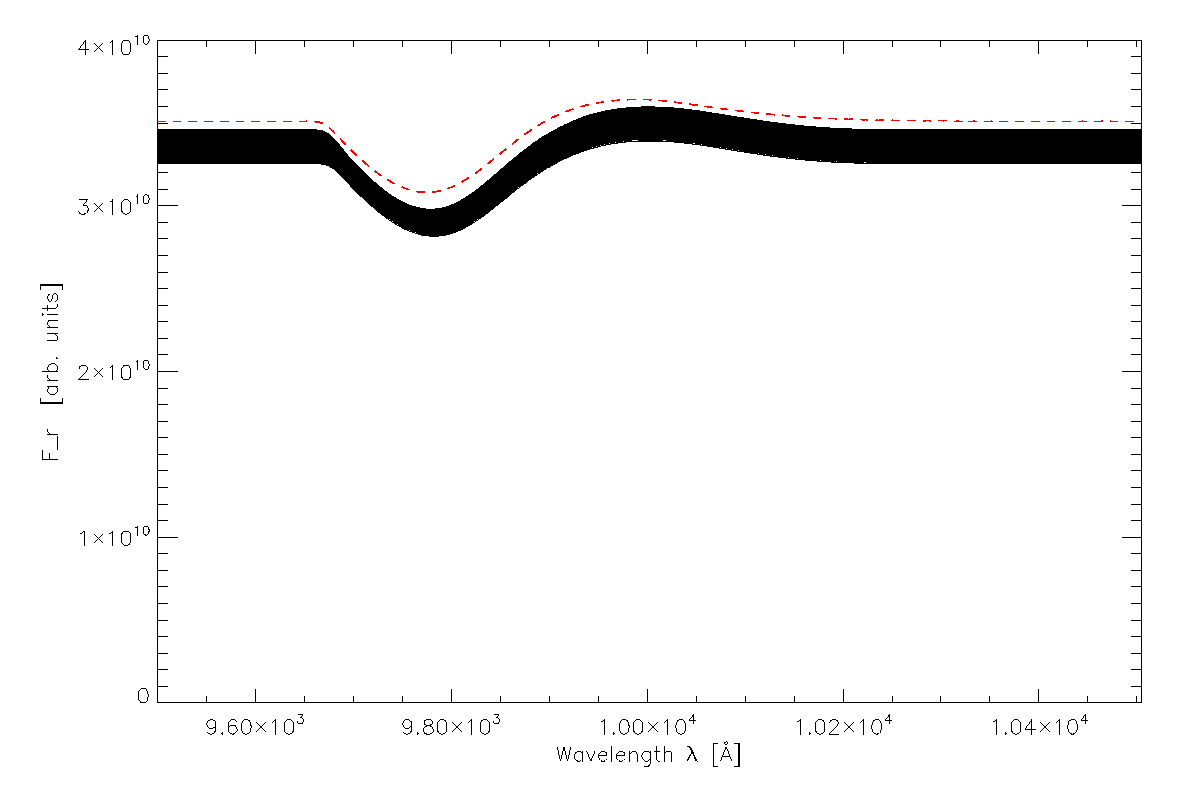} 
  \caption{Example of a calculation with a linearly increasing velocity field with $v_{max}=1\cdot10^4\ \kmps$. Solid: Spectra from all outermost voxels in the 3D spherical grid, dashed: Spectra from the 1D code. The Lorentz boost drives the continuum up as $v$ gets closer to $c$.}
  \label{v_near_c}
 \end{minipage}
\end{figure}

The formulation of the Eulerian method in the observers frame is inherently non-relativistic, which leads to differences between the 1D Lagrangian code and the new method presented here when the velocity field is greater then about $5\cdot10^3\ \kmps$. When the velocity is getting close to the speed of light, the Lorentz boost drives the continuum of the radiation field higher, what is clearly visible in Figure \ref{v_near_c}. As $v$ gets closer to $c$ this effect increases, the error in the Eulerian solutions increases rapidly.   

Various tests show that the agreement between the well-tested Lagrangian 1D and the Eulerian 3D code is excellent with velocity fields up to $1\cdot 10^3\ \kmps$, this is enough to do 3D radiative transfer in convection or global circulation models. We have extended the method described in \citet{CMF_five} into our 3D code, which then allows exact, full relativistic radiative transfer in 3D \citep[][E.~Baron et al., in preparation]{chen}  

%%%%%%%%%%%%% CONCLUSION
\section{Conclusion}
With our new Eulerian method it is now possible to do 3D radiative transfer in non-relativistic, arbitrary velocity fields in spherical, Cartesian, and cylindrical (while not described in detail, the method is also implemented in this part of the code) coordinates. The memory footprint and the computing time of the new algorithm in the two-level-atom setup presented here is negligible compared to the general requirements of the 3D code. In multi-level applications the time needed to calculate the opacity for every solid angle must be considered. The velocity field limitations must also be kept in mind when using the method.

%%%%%% acknowledgments
\begin{acknowledgements}
{ Some of the calculations presented here were performed at the H\"ochstleistungs Rechenzentrum Nord (HLRN); at the Hamburger Sternwarte Apple G5 and Delta Opteron clusters financially supported by the DFG and the State of Hamburg; and at the National Energy Research Supercomputer Center (NERSC), which is supported by the Office of Science of the U.S.\,Department of Energy under Contract No.\,DE-AC03-76SF00098. We thank all these institutions for a generous allocation of computer time. AS thanks the Research Training Group GrK 1351 of the German Research Foundation for funding. This work was supported in part by NSF grant AST-0707704 and by US DOE Award Number DE-FG02-07ER41517.
}

\end{acknowledgements}
\bibliography{3DRT_seelmann.bbl}

\end{document}